# Avoiding the Great Filter: A Simulation of Important Factors for Human Survival

Jonathan H. Jiang[1], Ruoxin Huang[2], Prithwis Das[3], Fuyang Feng[4], Philip E. Rosen[5], Derek Zuo[6], Rocky Gao[7], Kristen A. Fahy[1], Leopold Van Ijzendoorn[8]

[1.] Jet Propulsion Laboratory, California Institute of Technology, Pasadena, CA 91108, USA
[2.] Carnegie Mellon University, Pittsburgh, PA 15213, USA
[3.] Vivekananda Mission High School, Panskura, WB 721139, India
[4.] Beijing Normal University, Beijing, China
[5.] Independent researcher, WA 98662, USA
[6.] Sage Hill School, Newport, CA 92657, USA
[7.] University of California, Santa Barbara, CA 93106, USA
[8.] Department of Computer Sciences, Radboud University, the Netherlands

## Abstract

Humanity's path to avoiding extinction is a daunting and inevitable challenge which proves difficult to solve, partially due to the lack of data and evidence surrounding the concept. We aim to address this confusion by addressing the most dangerous threats to humanity, in hopes of providing a direction to approach this problem. Using a probabilistic model, we observed the effects of nuclear war, climate change, asteroid impacts, artificial intelligence and pandemics, which are the most harmful disasters in terms of their extent of destruction on the length of human survival. We consider the starting point of the predicted average number of survival years as the present calendar year. Nuclear war, when sampling from an artificial normal distribution, results in an average human survival time of 60 years into the future starting from the present, before a civilization-ending disaster. While climate change results in an average human survival time of 193 years, the simulation based on impact from asteroids results in an average of 1754 years. Since the risks from asteroid impacts could be considered to reside mostly in the far future, it can be concluded that nuclear war, climate change, and pandemics are presently the most prominent threats to humanity. Additionally, the danger from superiority of artificial intelligence over humans, although still somewhat abstract, is worthy of further study as its potential for impeding humankind's progress towards becoming a more advanced civilization cannot be confidently dismissed.

## 1. Introduction

Humanity's self-annihilation is a relatively new field of study as, apart from the extinction of our species having not come to pass, such capability was only acquired with the dawn of the nuclear age. Thus, to draw any conclusions from its study requires a substantial amount of theory, assumption, and estimated prediction. Given that the specter of self-annihilation looms over every aspect of our future, unflinching analysis leading to the means by which to avoid it is of paramount importance. As the eminent evolutionary biologist E.O. Wilson starkly stated, "The real problem of humanity is the following: We have Paleolithic emotions, medieval institutions, and godlike technology." In consideration, one might thus be forgiven if despairing to the point where termination by our own hand almost appears as an inevitable outcome of the very progress which has brought us to mastery of our world. One study [1] on the extinction of humanity concluded that almost all the risk comes from anthropogenic (human-made) causes. In other words, humanity's extinction will most likely be a consequence of our own actions. Another study [2] views self-annihilation as the answer to the Fermi Paradox [3] which, as posed in 1950 by physicist Enrico Fermi, asks the question of why there has been no contact with alien civilizations. Many possible causes of our self-annihilation, such as climate change and biotechnology [4], have been studied and modeled. However, due to the obvious lack of data regarding humanity's self-annihilation, assumptions vary widely across different studies and as such it is difficult to compare the effects of certain potential threats originating within our civilization relative to one another. As a result, what is popularly perceived as the most impactful and immediate threat to our survival often dominates debate and analytical speculation. Part of the challenge stems from the substantial influence human emotions and cognitive ability has over the likelihood of self-annihilation. These factors strongly influence human error, political decisions, and how we approach international relations, among the many other components which comprise the complex mosaic of the present-day world.

The probability of self-annihilation is a function of the potential threat which humans create during the act of harnessing great power and, once harnessed, the potential for humans to use such power in destructive ways. Thus, the event of self-annihilation is effectively determined by the outcome of a race between the technical knowledge humanity gains over time and our accumulated wisdom of how to constructively use such knowledge. A recent study [2] which attempts to solve the Fermi Paradox concluded that the probability of self-annihilation of complex life was the most influential parameter in the search for extraterrestrial intelligence. Another study [1], which predicts the likelihood of humanity's extinction, concluded that the annual probability of human extinction from natural causes is less than 1 in 87,000. Statistically speaking, this suggests humans would become extinct most likely through anthropogenic cause(s). Given the many ways in which humanity could bring about its own destruction, we first need to know which threats to prioritize. One study [4] highlights the dangers of proliferating and groundbreaking biotechnology, with humanity's survival time predicted to range between decades to centuries. The study concluded that due to the sheer growth in the number of individuals, institutions, and governments accessing biotechnology, this field poses a major threat in the near term. However, studies such as these do not reveal how dangerous certain threats are relative to each other. As previously mentioned, such questions arise from the lack of historical data, as well as the nature of human behavior and responses when confronted by existential threats.

Though not ignoring threats from asteroid impact and artificial intelligence, this study focuses on the chief contributors to self-annihilation of the human species. Through the measurement and analysis of both our technical knowledge and societal wisdom, the effects of nuclear war, climate change, asteroid impacts, pandemics and potential dangers from superior AI were simulated to determine estimations indicative of the time when humanity might expect to face the potential risks from such *Great Filter* events [34], in the near to distant future. Armed with such information, humanity could be better prepared to secure our long-term survival and that of the millions of other species which inhabit the Earth, while keeping in mind that the larger the remaining proportion of the Great Filter we face, the more conscious humanity has to become in terms of avoidance of negative scenarios [34] which might in fact pave the path for self-annihilation and would eventually inhibit our potential and aspirations towards attaining the status of a more advanced and "truly intelligent" civilization.

## 2. Methodology

For the study, we identified nuclear war, climate change, pandemics as well, as dangers from asteroid impacts and superior artificial intelligence, to be the most urgent and dominating scenarios in regard to the potential for self-annihilation. From a broader perspective, this study delves into predicted futures while simultaneously projecting the objective of estimated timelines for human survival on Earth (exclusive of the influence of any human expansion off-world) from simulations built using simple probabilistic models rooted in real-world scenarios. Enhancement of modeling is achieved by applying simple machine learning techniques. Given that we are (in part) attempting to find potential explanations to the seeming silence from supposed other technological civilizations in the cosmos, often referred to as the Fermi Paradox, we define a civilization to be destroyed when it loses enough population, resources, or technological capabilities that it no longer possesses the ability to communicate across interstellar distances such as would be required for signal contact with an extraterrestrial intelligence. We have used two slightly different types of simulations, this in terms of the underlying methodology, wherein the factors utilizing the former procedure (Simulation Type I) went through a simulation run of up to 100,000 epochs, each epoch covering as many years as it takes to destroy a civilization. The other set of factors, subjected to the latter class of simulations (Simulation Type II), have not undergone a run of 100,000 epochs, but are rather based on predictive analysis of machine learning models and mathematical modeling, as shown schematically in Figure 1.

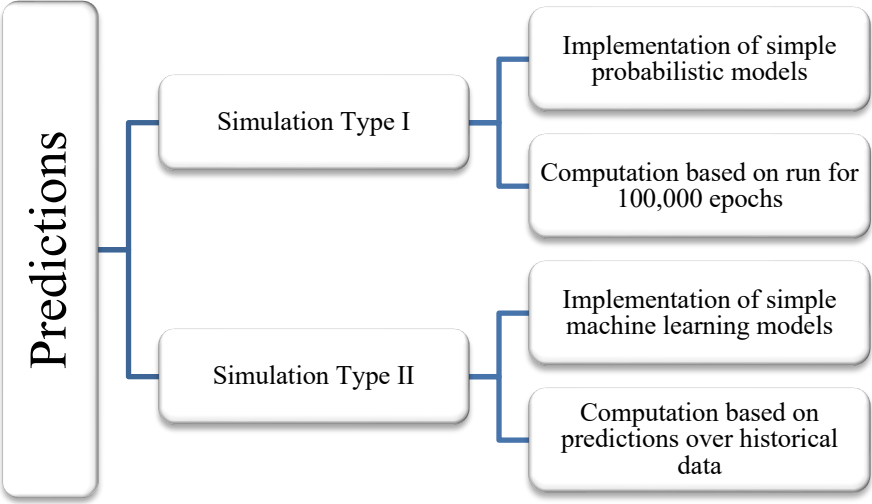

Figure 1: Classification of simulation models.

We chose 100,000 epochs after trialing over a range of values, finally settling on a large enough number to allow our simulation results to be stable - that is, the results fall within the range of a definite maximum as well as minimum.

## 3. Simulations

The simulations, based on the five major scenarios having the potential for deep and irreversible implications on the prolonged survival of human beings, encompass both the theoretical as well as computational aspects of our study. Every simulation is comprised of three critical components: assumptions applicable to a given simulation, the algorithmic and theoretical structure, and the final results based on computation. Well-substantiated data is at the heart of our predictive analysis, delivering a detailed overview of the past to the present, and by which the analysis attains a stronger basis with minimal errors - i.e., takes us closer to an accurate hypothesis about the future. Thus, the datasets, and in-turn the curated data, that constitute a major section of the simulations have been detailed under each section.

With respect to the simulations that we have employed while analyzing the important factors for human survival, we have abandoned the discussion of an otherwise helpful component of every simulation - the *uncertainties* in the estimations. The primary reason for such omissions is that there are not sufficiently well-defined conclusions, here owing to the algorithmic designs which tackle these real-world scenarios, as well as lack of relevant information reflecting the errors in the data. In setting aside errors which are inherent to the machine learning models being implemented for the future predictions, lack of quantified uncertainties defaults the results to a specific number of years after which we might expect to face the dangers from the five considered scenarios. In fact, the result may be less or more than the predicted number of years reported based on computation and the implications can indeed be alarming when the nearer-term bracketing case is considered. Finally, it must be kept in mind that the actual number of years for human survival is dependent on the unbiased superposition of all possible scenarios on a global scale. The guidance and plans we act upon in terms of preventive measures to safeguard human existence on Earth, as well as ensure a sustainable future, will ultimately determine our progress towards becoming a Type I civilization and beyond.

### 3.1. Nuclear War

For this simulation, we intend to use a function of the number of total nuclear weapons of the world (total TNT explosive equivalent) as a probability function in estimating the likelihood of human extinction by world-scale nuclear warfare.

### 3.1.1. Assumptions

Before we define the function of human extinction by nuclear war, we must make some pragmatic assumptions and limits:

- The probability $[P(N)]$ or the probability density function $[f(N)]$ can be estimated by the function of the number of nuclear weapons (i.e., total yield).
- $f(N)$ can be expanded into Taylor Series.
- $f(N)$ is defined as the growth rate of the probability of human extinction and is similar to the probability density function.
- When the number of nuclear weapons reaches a critical number, $N_0$, the probability density is 0, and the probability is 100%.
- When the number of nuclear weapons reaches another critical number, $N^*$, the density function, $f(N)$, will reach an inflection point.
- Supplementary condition: Being a probability density function, $f(N)$ shall be positive.

Considering all six assumptions, we have obtained the following four constraint conditions regarding $f(N)$

$$\int_0^{N_0} f(N)dN = 1 \qquad (1)$$
$$f(N_0) = 0$$
$$f(N^*) = 0$$
$$f'(N^*) = 0$$

Since we have four constraint conditions, the order of the Taylor Series will be four. Thus, we can attain the polynomial function of $f(N)$ as

$$f(N) = a_0 + a_1 N + a_2 N^2 + a_3 N^3 \qquad (2)$$

From the equations (1) and (2), we can derive a system of linear equations and by solving the following matrix equation, we can compute $f(N)$.

$$\begin{bmatrix} 12N_0 & 6N_0^2 & 4N_0^3 & 3N_0^4 \\ 1 & N_0 & N_0^2 & N_0^3 \\ 1 & N^* & N^{*2} & N^{*3} \\ 0 & 1 & 2N^* & 3N^{*2} \end{bmatrix} \begin{bmatrix} a_0 \\ a_1 \\ a_2 \\ a_3 \end{bmatrix} \begin{bmatrix} 12 \\ 0 \\ 0 \\ 0 \end{bmatrix} \qquad (3)$$

Furthermore, according to our assumptions, we can determine the probability function $P(N)$.

$$P(N) = \int_0^N f(N)dN \qquad (4)$$

In the following sections, we shall delve deeper into our prediction model and pose an in-depth discussion of the parameters being used.

### 3.1.2. The definition of $N_0$

In our prediction model, we choose the *Little Boy* (Hiroshima nuclear bomb) as a unit of measure to assess the probability. The Hiroshima bomb possessed the energy of about 10k-20k tons of TNT equivalent. For the simplicity of calculation, we stipulate it as a standard for 20kt. Next, we shall calculate the effective killing area of a standard 20kt nuclear bomb using two different methods of computation.

1) Considering c to be a constant and expressing the yield as multiples of 10kt, we may compute the ground range (effective killing radius) as

$$R_k = c \cdot (yield \ in \ multiple \ of \ 10kt)^{\frac{1}{3}} \qquad (5)$$

Implementation of the above formula and plugging in the required values, we obtain,

$$R_k = 1.493885 \cdot (2)^{\frac{1}{3}} = 1.88 \ km$$

Therefore, we can infer that the effective killing area $S_k$ of our previously defined standard nuclear bomb is

$$S_k = \pi \times (1.88)^2 = 11.1 \ km^2$$

2) We then utilize filtered data [22] to calculate the *ground range* of a 20kt nuclear bomb with a burst altitude of 540 m which corresponds to a *slant range* of 1.8 km (considering the total dose for acute radiation syndrome based on the effects of instant nuclear radiation). Assuming the direct radiation effects prevail at ground zero, the ground range can be computed using the Pythagorean Theorem as

$$R_k = \sqrt{SR^2 - h^2} \qquad (6)$$

Therefore, plugging in the values based on the data,

$$R_k = \sqrt{1.8^2 - 0.54^2} = 1.7 \ km$$

Thus, applying the second method, the effective killing area of a standard nuclear bomb is 9.26 $km^2$.

Combining the above two results, we stipulate the effective killing area of a standard nuclear bomb of 20kt is 10km$^2$.

$$S_k \approx 10 \text{ km}^2 \qquad (7)$$

According to the data provided by the World Bank and Knoema [23], the world's total urban and rural area accounts for,

$$S_{net} = S_{urban} + S_{rural} = 3,629,312 \text{ km}^2 + 47,953,427 \text{ km}^2 = 51,582,739 \text{ km}^2 \quad (8)$$

By division, we can compute N$_0$ as:

$$N_0 = \frac{S_{net}}{S_k} = \frac{51,582,739}{10} = 5,158,274 \qquad (9)$$

where $N_0$ is in the unit of the standard nuclear bomb.

### 3.1.3. The definition of $N^*$

Significant hemispherical attenuation of the solar radiation flux and subfreezing land temperatures may be caused by fine dust raised in high-yield nuclear surface bursts and by smoke from city and forest fires ignited by airbursts of all yield. Unlike most earlier studies, we find that a global nuclear war could have a major impact on climate manifested by significant surface darkening over many weeks, subfreezing land temperature persisting for up to several months, large perturbations in global circulation patterns, and dramatic changes in local weather and precipitation rates - a harsh 'nuclear winter' in any season [24].

For the second critical parameter, $N^*$, experts have created simulations based on several scenarios of nuclear exchange and predicted that there shall be a so-called "nuclear winter" after one yield of nuclear exchange. Although some researches have the opposite attitude towards the notion in terms of "nuclear winter" [25], they believe the concept of "nuclear winter" will have some problems. But for the sake of consistency, we hold a prudent attitude and believe that there will be a severe nuclear war aftermath when one yield of world-scale nuclear war breaks out between two or more nations.

In accordance to the study in [24], we consider six estimates of the total yield by summation of total nuclear warheads' yields: 100Mt, 500Mt, 1000Mt, 3000Mt, 5000Mt and 10,000Mt, as the critical parameter, $N^*$, of severe nuclear war aftermath.

### 3.1.4. The definition of growth rate — $t$

In this segment, we will define the growth rate of the world's arsenal of nuclear weapons. According to the data in [26], the trend of the global nuclear warhead inventories from 1945 to 2022 can be visualized from Figure 2.

Based on the literature data, we can calculate the average growth rate from 1945 to 1986, which constitutes the increasing part of the curve, and, for the decreasing part from the year 1986 to 2022.

$$t_{increasing-part} = 32.202\% \qquad (10)$$

$$t_{decreasing-part} = -4.623\%$$

In our model, we consider using a normal distribution to describe the growth rate as

$$t \sim N(\mu, \sigma^2) \qquad (11)$$

For the determination of the parameters $\mu$ and $\sigma$, we give two stipulations:

- $\mu = 0$: This consideration shall keep the random variable stable in general.

- $3\sigma$ Principle: It ensures that the sampling data has a very high confidence coefficient of 99.73%.

Therefore, we may conclude that

$$\mu = 0$$

$$\sigma = 4.623\%/3 = 0.0154$$

$$t \sim N(\mu, \sigma^2) = N(0, 23747 \times 10^{-4}) \tag{12}$$

Based on the computation in (12), we can argue that the nuclear weapons have no intentional creation or destruction in general. It ensures the randomness of the growth of nuclear weapons.

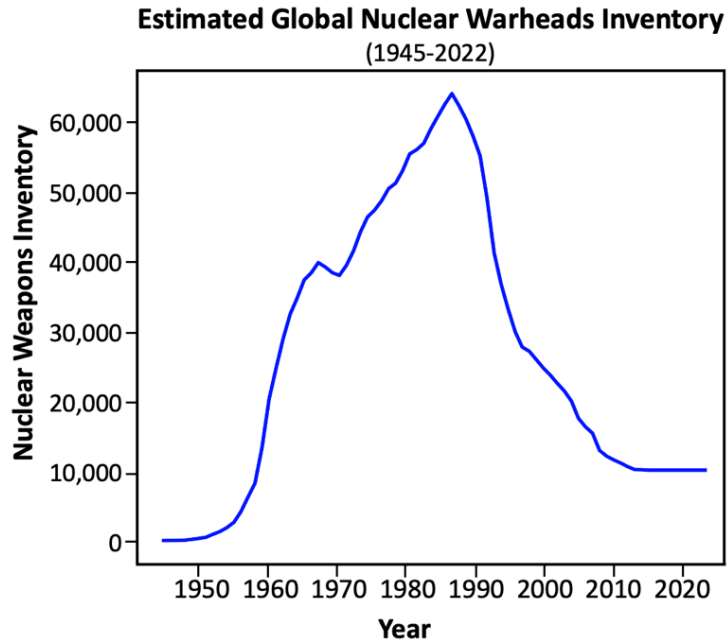

**Figure 2:** Global nuclear warhead inventories from 1945 to 2022.

With the help of (13), we can calculate the number of nuclear weapons in successive years. It is necessary to indicate that the number of nuclear warheads in 2022 is 12,705.

$$N_n = N_{n-1}(1 + 7), n \geq 1 \tag{13}$$

$$N(0) = 12,705$$

where $N_n$ is the estimated number of nuclear warheads $n$ years from the current year. An important point to note is that $N(0)$ represents the number of nuclear warheads in the year 2022 (0 is the means of considering 2022 as the reference year) and must not be confused with the first critical parameter $N_0$.

### 3.1.5. The procedure of simulation

Based on our computations so far, we can now determine the probability function $P(N)$. With the number of nuclear warheads for the reference year already obtained, we can compute a growth rate $t$ using random sampling from a normal distribution. By using (13), we can calculate the number of nuclear weapons in the years to come iteratively and then use random sampling from a uniform distribution to get a trial, and by the method of rejection sampling, determine how many years human civilization would last devoid of nuclear wars. A noteworthy fact is that we have set the upper limit of surviving years to be 20,000. Once a civilization last for 20,000 years, nominally chosen as the 'safe point' in the future beyond which any threat of nuclear war is presumed to be permanently in the past, the algorithm shall break out and run for the next epoch. We run this simulation for 100,000 epochs and obtain the average number of surviving years in a given scenario based on the algorithm in Figure 3.

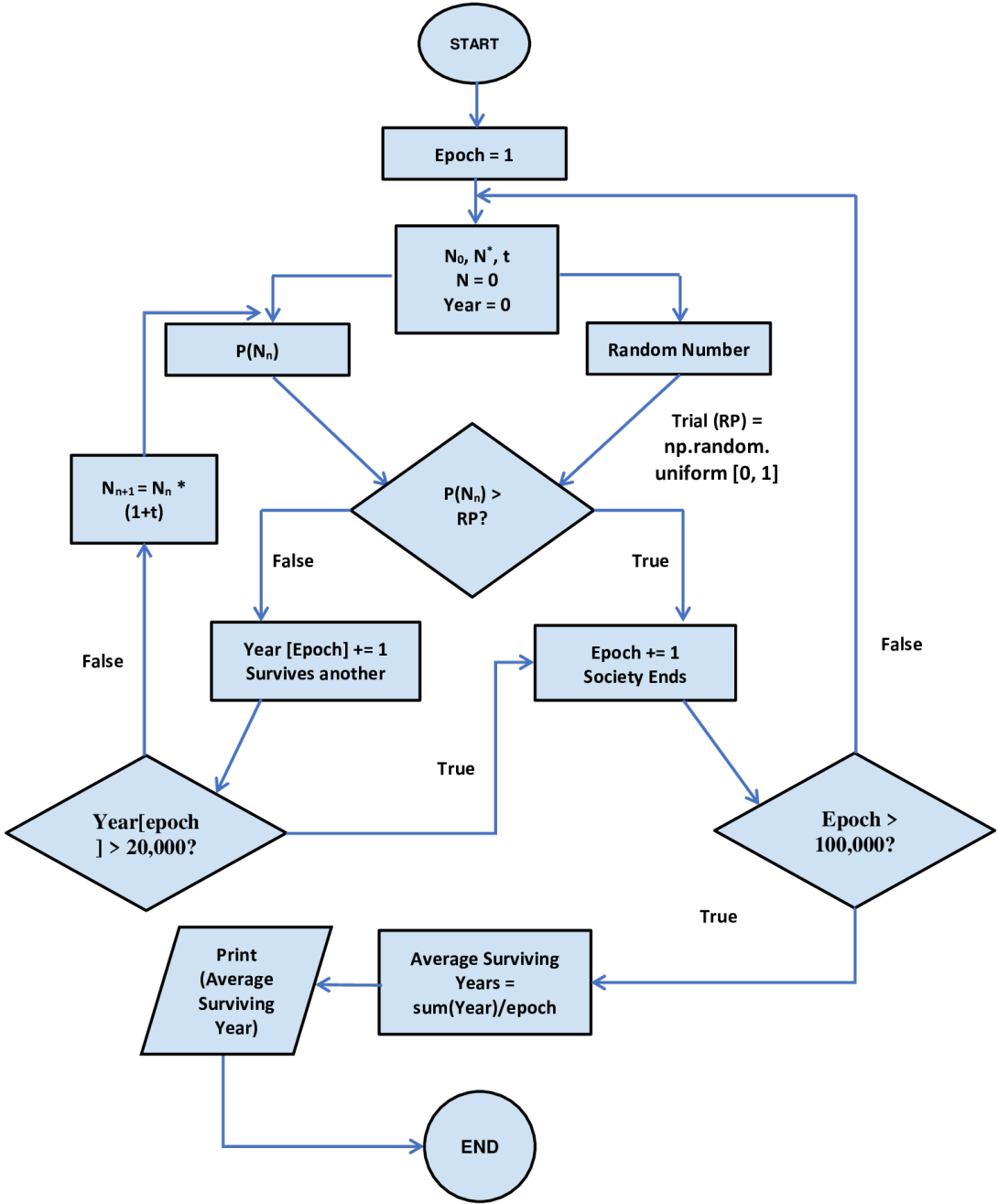

**Figure 3:** The program chart of the simulation.

### 3.1.6. Simulation results

Before we jump right into the results, a few facts must be taken into account. Firstly, we do not have information about the average TNT equivalent of our nuclear weapons and secondly, we are not certain as to how many of nuclear warheads would lead to so called "severe nuclear war aftermath". Thus, the results (listed in Table 1) are based on all types of average TNT equivalent and $N^*$. For example, when the average TNT equivalent is 1 million tons and $N^*$ is 100 million tons, the life expectancy of human civilization is only about 60 years.

**Table 1:** Average surviving years in different scenarios.

| Average TNT equivalent(tons) | Average surviving years in different $N^*$ | | | | | |
|---|---|---|---|---|---|---|
| | 10,000Mt | 5000Mt | 3000Mt | 1000Mt | 500Mt | 100Mt |
| 300,000 | 18,635.62 | 16,646.93 | 14,207.19 | 7717.29 | 6036.47 | 5000.80 |
| 400,000 | 17,857.17 | 14,844.43 | 11,181.57 | 4215.39 | 3048.76 | 2481.91 |
| 500,000 | 17,043.07 | 12,885.13 | 7936.82 | 2108.23 | 1496.07 | 1158.81 |
| 600,000 | 16,120.62 | 10,760.66 | 4825.39 | 972.26 | 697.4 | 539.55 |
| 700,000 | 15,135.78 | 8558.75 | 2588.03 | 454.59 | 340.9 | 264.11 |
| 800,000 | 14,048.18 | 6170.53 | 1251.27 | 225.14 | 168.49 | 144.46 |
| 900,000 | 13,001.1 | 3810.44 | 568.17 | 124.48 | 98.73 | 83.36 |
| 1,000,000 | 11,846.45 | 2050.4 | 262.29 | 72.99 | 62.79 | 58.44 |

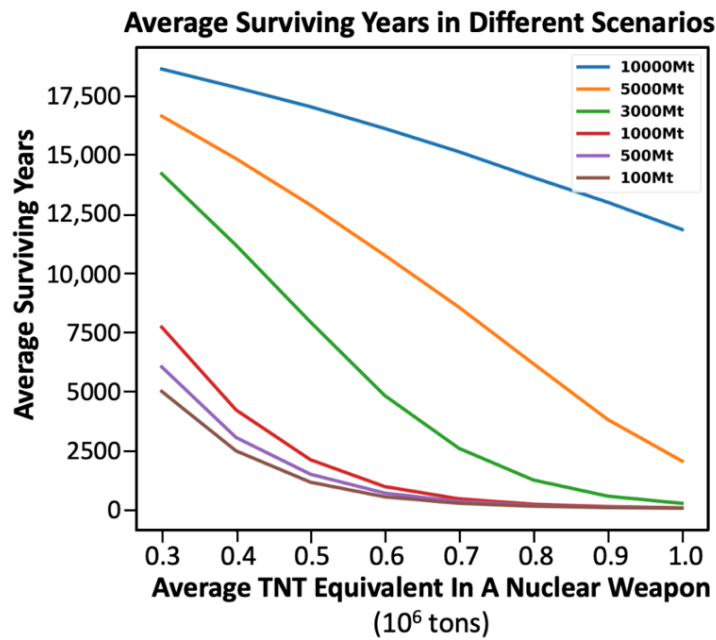

**Figure 4:** The figure of the simulation results.

In Figure 4, we plot the average surviving years for different values of $N^*$ and average TNT equivalent for each nuclear weapon based on the results in Table 1. Thus, we can infer that if the theory of nuclear winter is correct and our nuclear weapons arsenal is powerful enough, the life expectancy of humankind is indeed seriously threatened.

### 3.2. Climate Change

For the simulation investigating the threat from climate change, this primarily due to human activity, we determined the annual probability of civilization ending disaster scenarios. As, arguably, our warming world constitutes our most pressing concern for survival, we used global average temperature as an estimator for determining this scenario's probability. We chose to use a 2 °C rise in the average global temperature above the pre-industrial average to be the baseline tipping point. Coupled with the assumption that any year in which this change exceeds the 2 °C margin opens up a positive probability for a quickly cascading series of disastrous events leading to the destruction of civilization. Furthermore, we assume a rise of 12 °C above the pre-industrial average is a maximum critical level of temperature rise that will result in a 99% probability for globally disastrous events. This is chosen based on the fact that humans can only survive a maximum web-bulb temperature of 31°C (87 °F) at 100% humidity [28]. At a rise of 12 °C, the average temperature in many regions on Earth would pass the 35 °C threshold and become effectively inhabitable. We further assume the probability of disastrous events and the global average temperature to be linearly dependent – i.e., we assume a linear mapping from (2 °C, 12 °C) → (0%, 99%).

For our simulation we took into consideration the historical data comprising of the annual mean of the global land-ocean temperature from 1880 to 2021, provided by NASA's Goddard Institute for Space Studies [30], and applied non-linear regression to predict future temperatures (root mean squared error of 0.12) with no leveling-off or decline. As shown in Figure 5, the scatter plot in blue represents historical data and the red line, regressed to the historical data, is extrapolated accordingly into the future in absence of all effective mitigating factors.

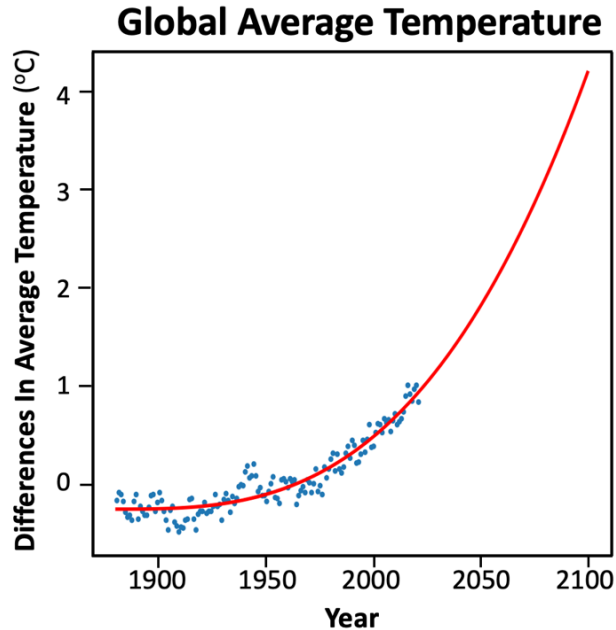

**Figure 5:** Historical data of differences in Global Average Temperature and Prediction curve ($R^2$ = 0.8977)

After running the simulation again for 100,000 epochs, we conclude the average time our society can survive before a civilization ending disaster prompted by climate change occurs is 193 years.

### 3.3. Asteroid Impacts

Asteroid and cometary impacts have played a major role in our Earth's history for eons. Approximately 65 million years ago, the seed of the human race's emergence was sown with the mass extinction of the dinosaurs, the cause of which is still evidenced by the Chicxulub crater located on the Yucatán Peninsula. But in accordance with the famous saying "*Those who do not remember the past are condemned to repeat it*" by George Santayana, this cataclysmic event can happen again and if so, will likely be the epilogue for our entire human race and that of a great many other species [6], [7]. In this section, we shall simulate and predict the average time the human race may survive before a massive asteroid impact- - another major challenge to avoid posed by the Great Filter. For this purpose, we have used the Sentry data, available from NASA's Center for Near Earth Object Studies (CNEOS), for analysis and it forms the basis of our simulation.

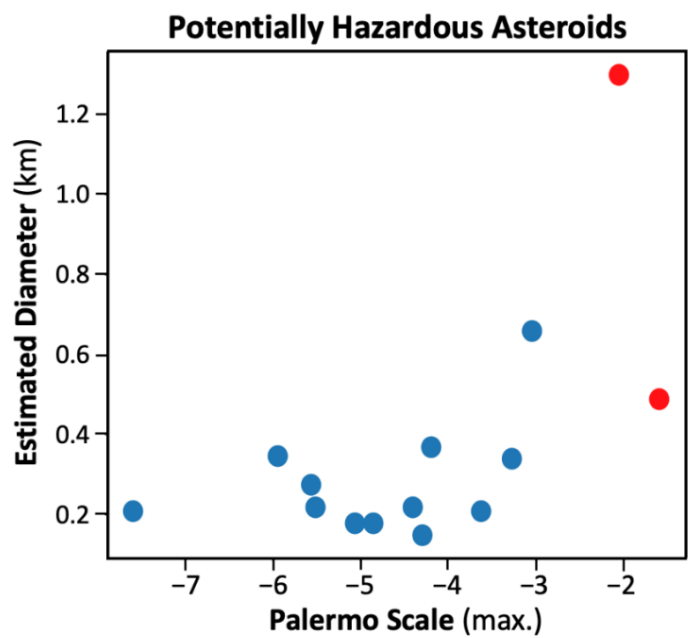

**Figure 6:** Plot of Estimated Diameters vs. Palermo Scale Readings of the PHAs.

We primarily focus on two attributes as a starting point for our simulation, namely the Estimated Diameter and Palermo Scale (a logarithmic scale used by astronomers in order to rate the potential impact hazard of a near earth object) readings. According to CNEOS, all asteroids which have an absolute magnitude (H) of 22 or less, and an estimated diameter greater than 140 m are classified as Potentially Hazardous Asteroids (PHAs). Imposing the above conditions on the CNEOS data brings the number of asteroids to 14, thus indicating the estimated number of PHAs. We additionally infer that the PHAs with a maximum Palermo Scale reading between −2 and 0 shall be considered for our simulation given it is those which have been determined to require careful monitoring.

From inspection of Figure 6, only two asteroids satisfy our condition, 101955 Bennu (1999 RQ36) and 29075 (1950 DA) which are indicated by the data points marked in red. Further, we have used the maximum of the impact probabilities for both of these asteroids as the constant annual asteroid impact probability and the prime input data for our algorithm in Figure 7.

Using the same general methodology as previously stated, the simulation was run for 100,000 epochs resulting in the average number of years of human civilization, devoid of any large-scale impact, to be 1754 years.

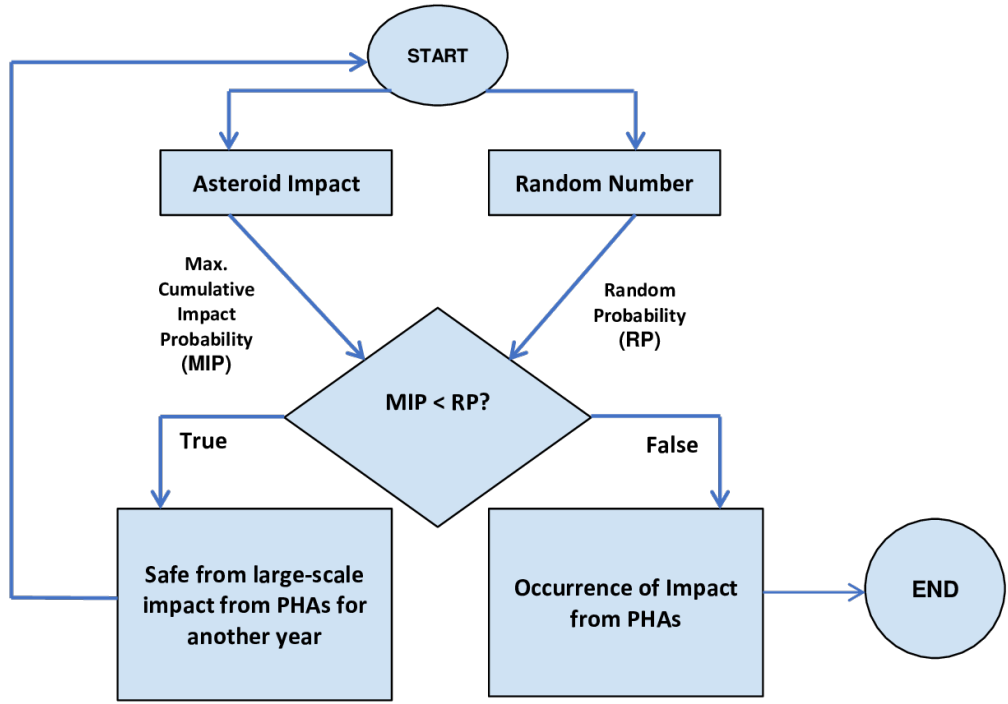

**Figure 7**: Simulation for impacts of potentially hazardous asteroids.

## 3.4. Artificial Intelligence

Artificial intelligence or AI is the ability of a computer or a robot controlled by a computer to do tasks that are usually done by humans requiring human intelligence and discernment [8]. The dawn of modern artificial intelligence was marked with the ideas of classical philosophers who tried to *describe human thinking as the mechanical manipulation of symbols*. Popularly advanced by the science fiction tales of Isaac Asimov and gradually turning into reality through subsequent developments, including the breakthrough study by Alan Turing on the possibility of creation of thinkable machines, the field of AI research has undergone rapid development. Examples can be found in dramatic breakthroughs in *Image recognition*, *Natural Language Processing*, autonomous robotics and in the world of gaming. Alongside data and algorithms, access to computing resources is critical for the advancement and diffusion of AI. While no widely used definition of AI compute capacity exists, it is comprised of a specialized stack of software and hardware (inclusive of processors, memory and networking) engineered to support AI-specific workloads or applications and includes supercomputers and large data centers [33]. Thus, there has been a special focus and projected exponential growth in the performance of supercomputers with their Rmax [a benchmark for supercomputer processors introduced by LINPACK, which is a software package for performing numerical linear algebra on digital computers] readings rising to a few hundred PetaFLOPS ($10^{15}$ Floating-Point Operations per Second). These advances have not only led to much improved scientific and engineering computation, but also to more advanced machine learning (ML)/AI models, this owing to immense capabilities in data manipulation and parallel processing. However, according to the Center for the Study of Existential Risks at the University of Cambridge, as AI systems become more and more powerful and much more generalized, they may exceed human performance. Passing such a threshold could lead to extremely positive developments - as well as giving rise to the chances of catastrophic risks in terms of safety and security which are now termed as the "dangers from AI". The concern is not with super-intelligence, but is with their unpredictable behavior to our requests [9] once they become superior to humans. According to Susan Schneider, director at the Center for Future Mind, "*It is simply the problem of how to control an AI that is vastly smarter than us.*" In this simulation, we intend to predict the danger from AI by attempting to compute the number of years before AI systems exceed human performance. For this purpose, we have analyzed the trend in FLOPS of supercomputers from 1993 to 2021 and compared that to the human brain equivalent of FLOPS to determine superiority, utilizing the algorithm depicted in Figure 8.

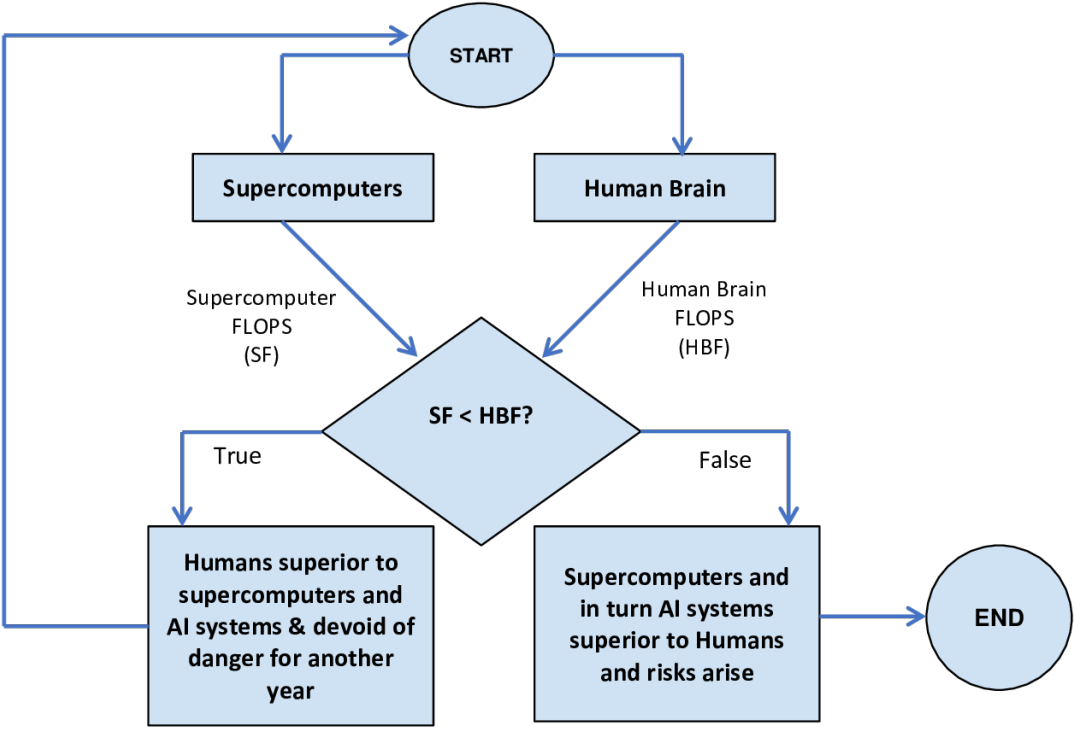

**Figure 8:** Simulation for estimation of years of superiority of human race over AI.

The reason for choosing the above methodology as the basis for our simulation is because *"We cannot consult actuarial statistics to assign small annual probabilities of catastrophe, as with asteroid strikes"*, according to Eliezer Yudkowsky from the Machine Learning Institute. Furthermore, since the advances of AI are significantly dependent on the *data* and the amount of *compute*, with more compute being a positive indicator towards the better performance of AI systems [32], it can be reasonably inferred that a comparison of the compute capacity, i.e., FLOPS of the world's fastest computing systems to that of an equivalent metric of the human brain is an effective way to simulate and predict the future scenario. To begin with, we studied the data on FLOPS of Supercomputers obtained from Top500.org (as in Figure 9) and applied linear regression for predictions of future FLOPS capability into the future.

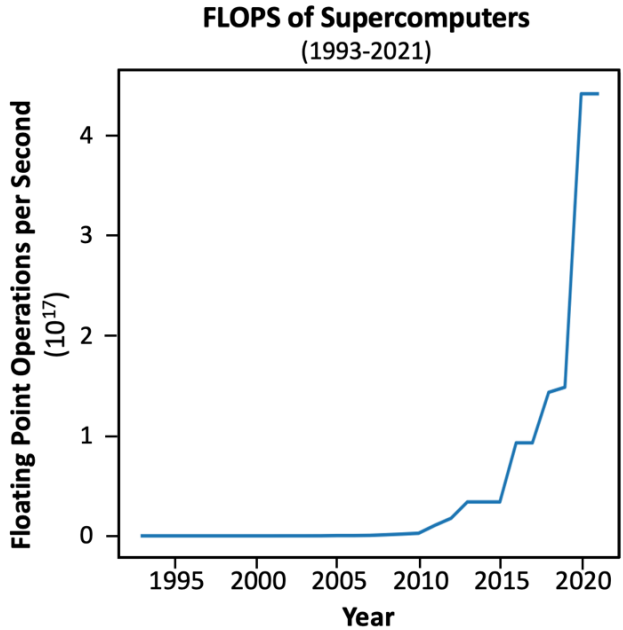

**Figure 9:** FLOPS of Supercomputers over the years 1993-2021.

Since application of linear regression requires an approximately linearized trend, we have modified our data to plot as log base 10 of the above, depicted in Figure 10.

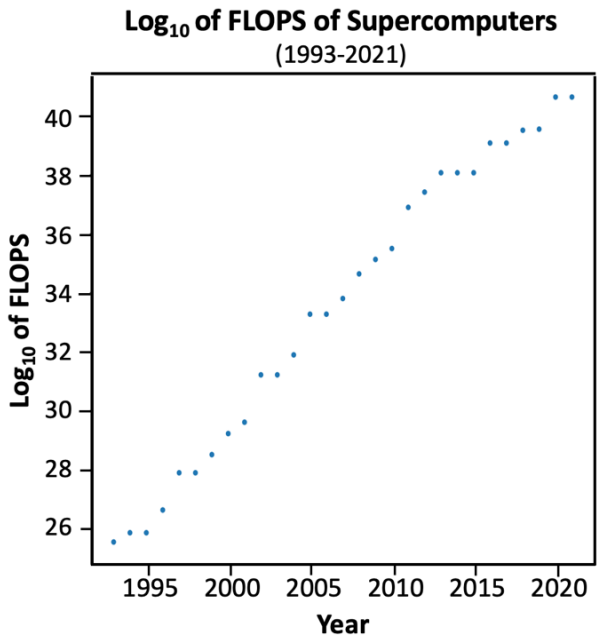

**Figure 10:** Log Base 10 of FLOPS of Supercomputers over the years 1993-2021.

An important fact worth noticing is that the trend shown by development of FLOPS of supercomputers has a striking similarity with the trend shown by the transistor count per microprocessor over the years 1993-2021 as well as with that trend's future prediction based on Moore's Law. This correlation is expected given transistor count strongly affecting the performance of supercomputers. Finally, in order to obtain the results of our simulation, we require the FLOPS of the human brain, which can be obtained from the Sandberg and Bostrom Project of 2008 [10]. Here we have used the estimate computed by Tuszynski (2006) for level 10 emulation which incorporates the *molecule positions, or a molecular mechanics model of the entire brain* in addition to the factors considered in the other levels of the hierarchy. Therefore, based on our simulation results, the prediction curve (Figure 11) suggests supercomputers, and in turn AI systems, will become superior to humans in 40 years, beginning from the current year.

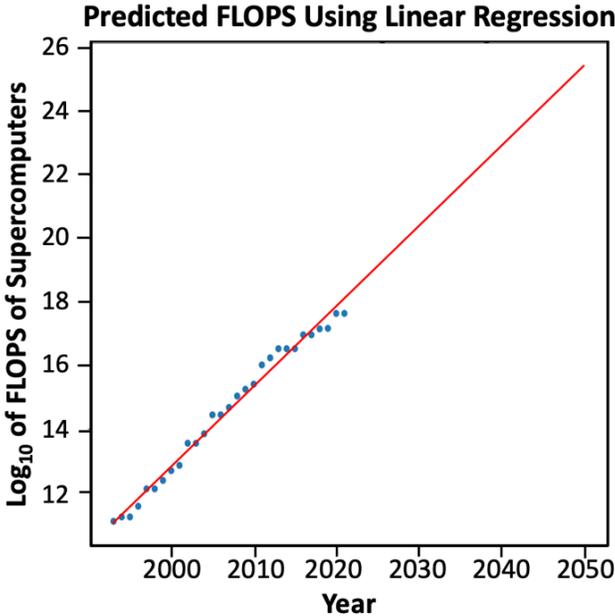

**Figure 11:** Plot of Predicted Results (including historical data) using Linear Regression with a *Root Mean Squared Error* of 0.24.

But predictions do not always match the reality. According to Neils Bohr, "*Prediction is very difficult, especially if it's about the future*". Further analysis of the trend in performance development of supercomputers over time implies (based on the top-ranked per Top500.org) they are gradually approaching an optimal performance limitation. Careful visualization of the section of the graph consisting of the sum of FLOPS of all the top 500 supercomputers over the years, obtained from the *Projected Performance in contrast to actual performance* plot over time by Top500.org, implies a maximum performance in the range of 1 to 10 ExaFLOPS. This is mainly due to the hardware limitations owing to our current level of technological advancement in materials and manufacturing. A notable instance of this is the increase in the performance (in terms of Rmax) of Fugaku, which has not seen the slightest increase in its performance since November of 2020 and only a "modest increase" [12] from June of 2020. Additionally, according to researchers at Top500.org, *performance growth rate used to be 1000x in 11 years, but now it is 1000x in 20 years*. This slowing of growth rate in performance is due to the fact that *Moore's Law is facing significant technological barriers and it is becoming progressively more difficult to increase the capabilities of processors at historical pace.* From the point of view of artificial intelligence, the HPL-AI benchmark (although presently discontinued, sought to highlight the emerging convergence of high-performance computing (HPC) and AI workloads) readings of Fugaku have gradually reached a halting point, a constant value of 2.000 Eflop/s as depicted by Figure 12.

Furthermore, based on the High Performance Conjugate Gradient (HPCG) benchmark, which provides an alternative to the benchmark posed by HPL, the highest estimate is about 16,004.50 HPCG-TeraFLOPS and has been achieved by Fugaku as per the HPCG results of November, 2021 and June, 2022. Thus, compared to the FLOPS of level 10 emulation of the human brain, which accounts for an estimate of $10^{16}$ TeraFLOPS, the highest HPCG estimates of the top-performing supercomputers over the years from 2017 are still far from achieving the capabilities of the human brain.

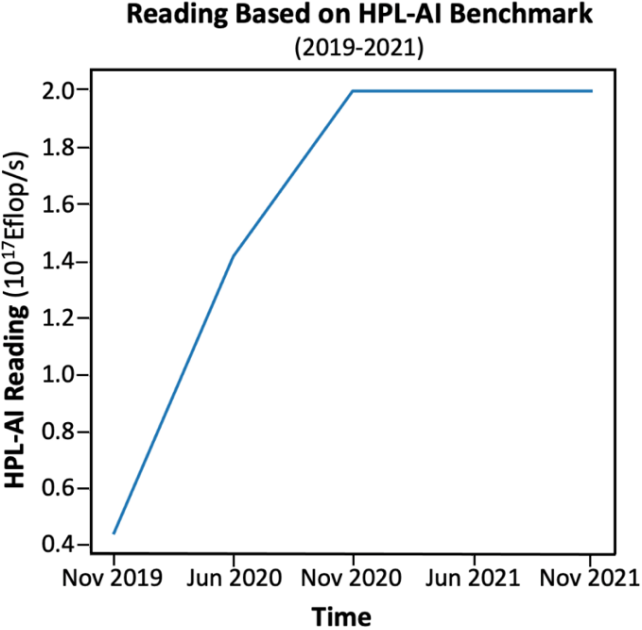

**Figure 12:** HPL-AI readings from 2019-2021 (as per available data).

Another aspect of our simulation is the fact that performance benchmarks like FLOPS largely measure how fast a computer can perform individual computations. But it does neglect the fact that supercomputers also need to move information across various components of itself, which takes up time, space and wiring, and ultimately affects their overall performance. To overcome this challenge, we use the Transverse Edges per Second (TEPS) metric for this and for the purpose of measuring performance of the human brain. Considering the upper bound of the estimate of human brain TEPS (about $6.4 \times 10^{14}$ TEPS) [11], we can imply that the brain performs 640 trillion TEPS and for computer hardware to reach that level of performance, the cost would range up to more than 1.489 billion US dollars per year. The current cost of Fugaku is about 1.213 billion US dollars - but still is not at the performance level of a human brain's communication system. Therefore, it is much more likely that although supercomputers, and their AI system cousins, may become technically superior to humans in terms of computation performance and TEPS in the next 40 years, it will still be far from reaching fully realized human level intelligence, especially because of the ability of the human brain to adapt and analyze changing situations and imagine the future. In contrast, AI heavily relies on past data which does not necessarily upgrade it to the mark of human level intelligence and/or senses – at least within the limits of current technology.

In the present scenario, researchers have shed light on the vast possibilities of quantum computing (QC) which have the potential to eliminate the obstacles to Artificial General Intelligence (AGI), as well as remove the technological barriers to effective continuation of Moore's Law. *AGI is the hypothetical ability of an intelligent agent to understand or learn an intellectual task that human beings can* [13]. Achieving AGI, with the aid of QC, would upgrade AI systems to the level of intellect of the human brain. But the fact that QC is still under active research and development rules out realizing this possibility, at least in the very near future. Accordingly, there is still much research and understanding to be gained in the internal working of computational technology in order to create more advanced AI systems. Mission critical as well is the research and development of technology necessary to control these systems effectively, and in so doing help to pave the way for human civilization to reach Kardashev Type I status and beyond [27].

### 3.5 Pandemics

Besides analyzing the possibilities of nuclear war or predicting the timeline for potential climate disaster, there still remains an ever-concerning fact that we must take into account while predicting the timeline for human survival on Earth - *infectious diseases.* As competing forms of life on Earth, pathogens have always been an integral part of our lives. With the rapid advancement of medical science leading to in-depth study of the pathogens, we are now well equipped to combat many diseases with effective counter-measures implemented by governments and/or private foundations, ensuring public health and safety. According to the Mayo Clinic, *infectious diseases are disorders caused by organisms like bacteria, fungi, viruses and parasites*. Interestingly, many of these live in and on our bodies and are normally harmless; but under certain conditions, some may cause diseases as severe as HIV/AIDS which claimed the lives of more than 36 million people worldwide, according to the World Health Organization. Some infections are more contagious than others, spreading in airborne form from one person to other (e.g., the common cold virus) while others can spread from insects, birds, animals and consumption of contaminated food or water. With the creation of efficiently distributed vaccines, diseases like chickenpox, smallpox and measles can now be prevented from posing threats to the lives of patients. But tracing back to the basic infrastructure of our medical research and healthcare systems, we must confront an unavoidable question: are we prepared for a new, and perhaps, highly infectious disease for which we may not have any prior experience or preventive measures? According to the World Health Organization (WHO), in the past 50 years scientists have identified more than 1,500 new pathogens; most of them began from animals and then spread to humans - i.e., zoonotic transmission [14]. In this simulation we attempt to predict a potential pandemic, including its timing for emergence in the future.

Before we delve deeper into the analysis, we must consider certain morbidity measures related to epidemiological studies which form the basis of virtually every model predicting possible future pandemics. These include the basic reproduction number ($R_0$) and the attack rate, which will be of paramount importance for this analysis. First introduced by Sir R.A. Ross in 1911 and later implemented by the Kermack-McKendrick compartmental model for prediction of epidemics, the basic reproduction number is defined as the *average number of infections caused by a single infected individual introduced into a wholly susceptible population over the duration of the infection of this individual* [18]. The Kermack-McKendrick model pointed out an important fundamental property of $R_0$, which is stated that in general, the disease outbreak will not develop into an epidemic if its $R_0$ value is less than 1, but will do so if the value exceeds 1 (note: we will be using this property in the later part of this section.) The attack rate - also termed as incidence proportion or risk - is defined as the number of new cases to the number of people at risk in the population. In other words, it implies the probability of a person developing the disease under consideration [21].

As the starting point of our simulation, we have used an assessment framework [15], which comprises of a four-step process, to create a mapping between basic reproduction numbers ($R_0$) of flu viruses and the scaled measures of transmissibility. The framework includes identification and evaluation of transmissibility and severity, followed by scaled measures of both based on literature review of past pandemics, summarization of the scores and measures (as shown in Table 1), and finally historical context which considers the four major pandemics of the last 100 year (in 2009, 1968, 1957 and 1918) and three non-pandemic influenza seasons (in 1978-79, 2006-07 and 2007-08.) A refined assessment is created which incorporates finer scale allowing more discrete separation of pandemics and flu-seasons, in addition to more epidemiological and clinical information. In our methodology, we have primarily utilized the first three steps where the range of measure for transmissibility was divided into a five-point scale as shown in Table 2, below.

**Table 2:** Scaled measurements of transmissibility on the basis of $R_0$

| Transmissibility | Scale | | | | |
|---|---|---|---|---|---|
| | 1 | 2 | 3 | 4 | 5 |
| Basic Reproduction Number | ≤ 1.1 | 1.2-1.3 | 1.4-1.5 | 1.6-1.7 | ≥ 1.8 |

The reason for choosing influenza pandemics as the baseline of our analysis is that pandemic influenza has occurred at fairly regular intervals throughout reliably recorded history and will be repeated [16]. This is mainly because pandemic influenza strains can cause severe infections that often are accompanied by non-insignificant fatality rates [17]. To analyze the severity of a pandemic, we need to compute the attack ratio (usually known as attack rate), which is widely used for hypothetical predictions and for proper estimates during actual outbreaks, defined by the equation:

$$A = 1 - \frac{S_f}{N} \tag{14}$$

where $S_f$ is the number of susceptible individuals at the end of the pandemic and N is the total population, assuming to be constant [18]. A mathematical relationship establishes relation of the basic reproduction number to the attack ratio according to the equation [18]:

$$\ln\left(\frac{S_0}{S_f}\right) = R_0\left(1 - \frac{S_f}{N}\right) \tag{15}$$

where $S_0$ is the number of susceptible individuals at the *start of the pandemic* which we can approximately consider to be equivalent to the entire population, N. Thus, (15) can be utilized to analyze the attack rates based on the reproduction numbers of the previous pandemics computed from historical data. Substituting N for $S_0$ and combining (14) and (15) gives the following result:

$$(1 - A)e^{AR_0} = 1 \tag{16}$$

Addressing our primary assumption involved in the derivation of (16), we have analyzed the data based on the first wave of the pandemics since the very first wave typically marks the period of highest susceptibility for the entire world population due to the presence of no effective vaccines or natural immunity. As the input data for our simulation, we have used the minimum of the basic reproduction of the *first wave* of the 1918 influenza pandemic (of those which occurred in the era of modern virology and were declared as epidemics) which was essentially 1.03 with a 95% CI of 1.6 to 2.0 [19] and is being considered due to its reputation as the most severe influenza pandemic the world has ever known.

One important reason for excluding the ongoing pandemic widely known as COVID-19, which is caused by the SARS-CoV-2 virus, from the scenario is the fact that although both COVID-19 and influenza are contagious respiratory illnesses and have similarities in their symptoms, they are caused by two different viruses [20]. Attack rates play a key role in our simulation (and in fact, other epidemiological models as well) since they facilitate the path for their comparison to the random number probabilities, a procedure that we have implemented for the other human survival simulations in order to obtain prediction results.

After running our simulation for 100,000 epochs based on the algorithm in Figure 13, our calculations indicate the possibility of an outbreak after 16 years (starting from the current year) which will eventually turn into a *growing influenza epidemic* owing to its minimum basic reproduction number being greater than 1. An important point to consider is that the above result does not guarantee that the predicted epidemic will definitely turn into a pandemic, but the potential exists if effective countermeasures are not taken in a timely manner.

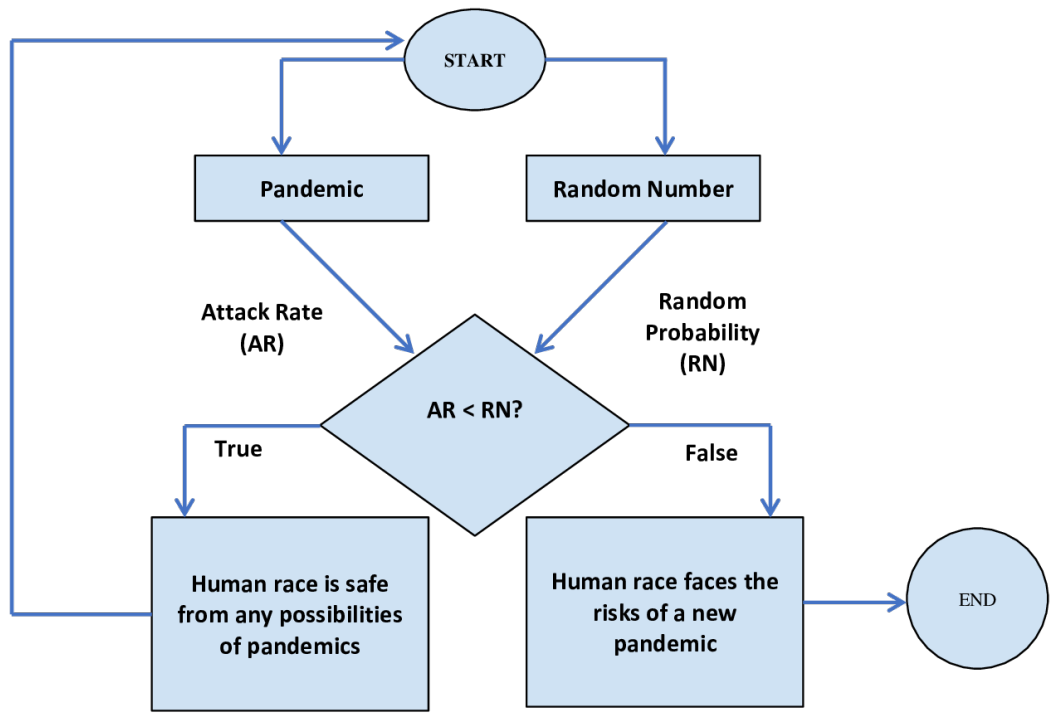

**Figure 13:** Simulation for predicting potential pandemic in the future (repeated 100,000 epochs).

Other than its direct impact human lives, future pandemics also have the potential to inflict great amounts of damage to our civilization from rising political tensions, economical disasters, emotional crises, social upheavals, degradation of educational institutions, etc. For example, during 2021 the COVID-19 pandemic, according to the Global Peace Index, contributed to a 10% increase in civil unrest, 60% of the global population is worried about sustaining serious harm from violent crime, and the economic impact of violence increased in 2020 to $14.96 trillion. Moreover, during the pandemic the world's economy also took a huge hit with the median global GDP dropped by 3.9% from 2019 to 2020 according to estimations by the International Monetary

Fund (IMF), this as countries directed resources to respond to the pandemic while other important areas of were neglected. Another area of impact is the social implications of pandemics – e.g., COVID-19 has changed how we communicate, learn, work, and how we see the world. By predicting how future pandemics could happen, we not only foresee how we could be affected, but also how we could change as a society. COVID-19, in terms of fatality rate alone when compared to other major pandemics of human history, is relatively mild. If a pathogen of the magnitude of the Bubonic Plague or the Spanish Flu was able to spread in today's hyper-interconnected world, the consequences could be far more disastrous.

As a society, we have to responsibility to prepare for such pandemics and respond to them accordingly; however, different ideologies often prevent us from working together, putting at risk humanity's future. Whether it is over simple disagreements such as wearing a mask or getting vaccinated, or how different countries handle the pandemics on an institutional scale, humans are often incapable of acting in concert. If we can't unite to effectively direct resources to important sectors for our survival instead of investing in ever larger militaries, we might not survive another 100 years. Pandemics have perhaps the most complex effects on society among the threats to human survival on Earth explored in this paper. Conversely, unlike asteroid impacts or nuclear war, once a pandemic happens we still have a chance to combat it using measures such as masks, social distancing, vaccines, limiting travel, and more. As COVID-19 has demonstrated, pandemics, when they arise, must be reckoned with and can be either mitigated or exacerbated by human behavior - e.g., wearing masks that work to protect everyone, to the inequality in distribution of lifesaving vaccines between poor and rich countries, human behavior is a contributing cause as to why the current pandemic has gone on for so long. There is an age old saying that "*an ounce of prevention is better than a pound of cure.*" If we are to prevent epidemics (and in-turn pandemics) from claiming the lives of millions of innocent people, we need to take the critical steps of ensuring strong leadership and global cooperation, strengthening healthcare facilities in poor countries, creation of a Global Epidemic Response and Mobilization (GERM) team which will always be prepared to mount a well-coordinated response to the next threat of a pandemic [14] and above all, major improvements in medical technology and healthcare infrastructure for the creation and implementation of the most effective vaccines in the least possible timespan. These steps, yet difficult to attain, will not only help pave our path to avoiding the Great Filter.

## 4. Summary of Results

The results derived from all the simulations that reflect the key factors affecting the survival of the entire human race are summarized in Table 4.

The nuclear war simulation models probability of human extinction using a polynomial function. By defining two key parameters, and, and the increase rate, we derive the analytical expression of the probability function. In calculating, we use an averaged value to improve accuracy while prudently leveraging two methods. For, we refer to the results by Turco et al. (1983) [24]. For derivation of the increase rate, the basis is a normal distribution coupled with the Principle. Then, by setting different average explosive yields in terms of megatonnage of TNT equivalent and, we can attain the simulation results - average surviving years. We have set the upper limit of survival interval to 20,000 years, rationally noting that if civilization lasts that long the threat of nuclear war will have been effectively relegated to the past. A consequence of this approach is that if the average TNT equivalent is relatively low and the $N^*$ (the critical number of total yield in triggering nuclear winter) is very large, the real average surviving years will be larger than the simulation results. However, this artifact of the calculations can be ignored as it is effectively disconnected to the simulation's objective - the minimum survival expectancy. For example, when the average TNT equivalent is 1,000,000 tons and equals to 1MT, the model indicates our civilization will not survive beyond 60 years. Since the height of the Cold War the number of nuclear weapons in the world has indeed declined, tacitly indicating a period of nuclear peace for current society. However, in the event of a large enough nuclear exchange our simulation results show that human society will not last very long when considering the consequences of "nuclear winter" theory. This sobering conclusion carries important implications for the world's current leadership. Whether the total number of nuclear weapons in the world is enough to destroy human society once, or even many times, "once is enough." Further, redundancies in nuclear weapons are not necessarily better for national security as it is predictable too many such weapons will increase the level of tension in international relations, at least to some extent, which is tantamount to lowering the threshold for their use. In recent years this has been exemplified by the "North Korean nuclear issue". It must be pointed out, however, 77 years have elapsed since the bombings of Hiroshima and Nagasaki. Thankfully in that time all nuclear capable nations have refrained from using these most lethal weapons in any conflict with other nations. We have only one human society, and likely only one chance at evolving to a higher technological civilization (e.g., at least Kardashev Type I [27]), so how to survive and thrive is our common fundamental task since entering the nuclear age. The premise of development is the continued existence of a supportive ecosystem to serve as a foundation. Whether by divine intent or the cold nature of an indifferent Universe, we cannot assume we are afforded a chance to simply start over if we allow civilization to slip from our grasp, such as depicted in Cixin Liu's landmark novel, The Three-Body Problem. In the final analysis, reality offers humanity no "restart" button so we must take care to preserve and grow what we, as a species, have come so far to build.

**Table 4:** Simulation results for all parameters.

| Simulation Type | Result (years) |
|---|---|
| Nuclear War | 60 (over 100,000 epochs) |
| Climate Change | 193 (over 100,000 epochs) |
| Asteroid Impact | 1754 (over 100,000 epochs) |
| Artificial Intelligence | 40 |
| Pandemic | 16 (over 100,000 epochs) |

The Climate Simulation takes into account the average global temperature trend and transforms that into an extinction threat probability. The main assumption for the simulation is that the apocalyptic scenarios are being considered once we hit the mark where the rise in global average temperature reaches 2 °C, and scales linearly thereafter to 99% probability of extinction at 12°C. Based on our prediction model (Figure 7), a rise of 2 °C in global average temperature would likely occur in 2054, i.e., 32 years from the present year. Climate change would end the human race slower relative to nuclear wars, primarily due to the slow rise of global average temperature and its complex consequences taking some amount of time to build to significant actual extinction circumstances once that threshold is reached. From the distribution illustrated by Figure 14, in this scenario most runs of the simulation fail to survive beyond 200 years, thus alerting us once again to what is perhaps the most pressing threat since the beginning industrial revolution.

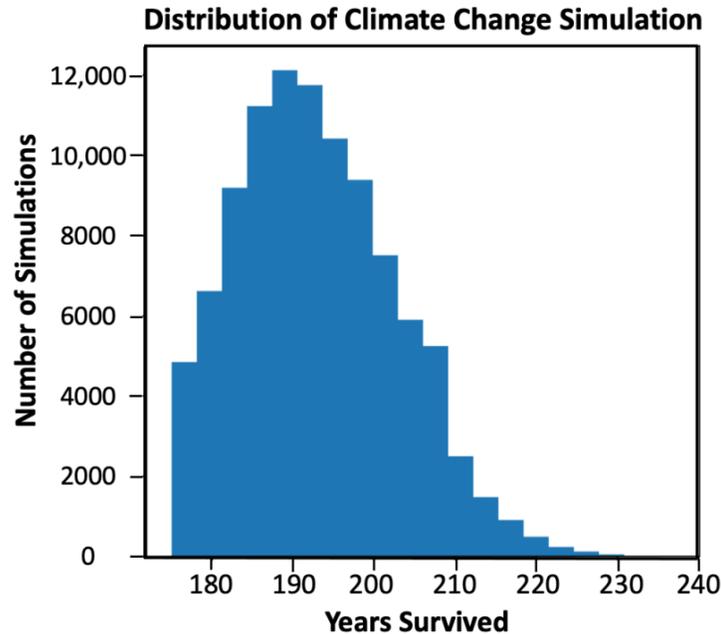

**Figure 14:** Distribution plot for Climate Change Simulation.

Consideration of the potential danger from asteroid impacts, the simulation model deals with the probabilities posed by *potentially hazardous asteroids* determined on the basis of their estimated diameters and absolute magnitudes. From the pool of 17 potentially hazardous asteroids, filtered out from the Sentry data of near-Earth objects by JPL, we have chosen the asteroids that have a maximum Palermo scale reading of greater or equal to -2 (since these are the NEOs requiring special attention, as per the official documentation), thus paring down the number of PHAs under consideration to just two. Furthermore, the simulation utilizes the maximum impact probability of these two due to the fact that any random probability greater than the maximum would account for a simulated destruction of the human race during each run of the algorithm. Analysis of the distribution plot in Figure 15 poses a clear revelation that most runs of the simulation fail to find civilization surviving after 2500 years have elapsed, yet the threats from asteroids are much less significant than other factors considered in this study. This suggests further attention is needed to the underlying assumptions, modeling and analysis of generated probabilities.

The artificial intelligence simulation, unlike the others in this study, utilizes a different approach towards the computation of prediction results. In terms of the extent to which a factor under consideration affects human survival altogether, the scenario for AI is relatively complex owing to its deep societal implications leading to both the benefits and dangers this technology might pose to the human society in the near future. Therefore, the approach of dealing with probabilities in order to compute the average number of survival years over 100,000 epochs is not a feasible option, primarily due to the lack of "actuarial statistics" to assign component probabilities, just as was the case with other important factors in the simulation. A solution to this problem derives from what lies at the heart of AI, i.e., increase in the computing power, owing to the fact that the advancements of AI systems is directly proportional to the fast and efficient processing of vast amounts of data. Thus, for the purpose of this simulation, we have chosen Floating Point Operations per second, or FLOPS, a metric of measuring the performance of supercomputers, which are the fastest computers ever built for bit-based computing. The big picture of the simulation is based on the algorithm of comparing the FLOPS of the human brain to that of the supercomputers based on future predictions using machine learning. This strategy would in turn signify the fact that once the performance of supercomputers exceeds that of the human brain, we must acknowledge the superiority of the former. This would, in fact, lead to the superiority of AI systems due to their higher capabilities than humans in terms of computational capability, thus moving much closer to AGI. In this idealized scenario, the AI systems are likely to become more advanced than humankind in the next 40 years. But in reality, this is not necessarily an appropriate conclusion to be acted upon. Since 2005, there has been a deceleration in the performance of supercomputers (as a sum of the performances) due to the inability of upgradation of processors at the rapid pace of previous developments. A reasonable explanation in terms of quantum mechanics is that the Heisenberg's Uncertainty principle defines a (seemingly) hard limit to the miniaturization of critical microchips themselves. As stated by Dr. Moore himself, if electrons are considered to be the smallest possible transistor components, then the year 2036 would be a reasonable prediction, signifying the time of convergence of Moore's Law and quantum physics [29]. Therefore, the slowing ascent of performance shall indeed make the superiority of AI systems a fact of the distant future, if at all. Furthermore, in terms of Transverse Edges per Second estimates, which additionally incorporate communication capabilities, computing systems are far from nearing the performance of the human brain. From a slightly different aspect, the immense power of imagination of humans with our unique ability of expressing emotions has always proved the unique factor and not likely to be matched by any creation of humans. Nonetheless, dedicated efforts to achieve this level of excellence in terms of AI systems has led to a much deeper understanding of how the brain functions, helping to unravel some of the mysteries of this fantastic creation of nature.

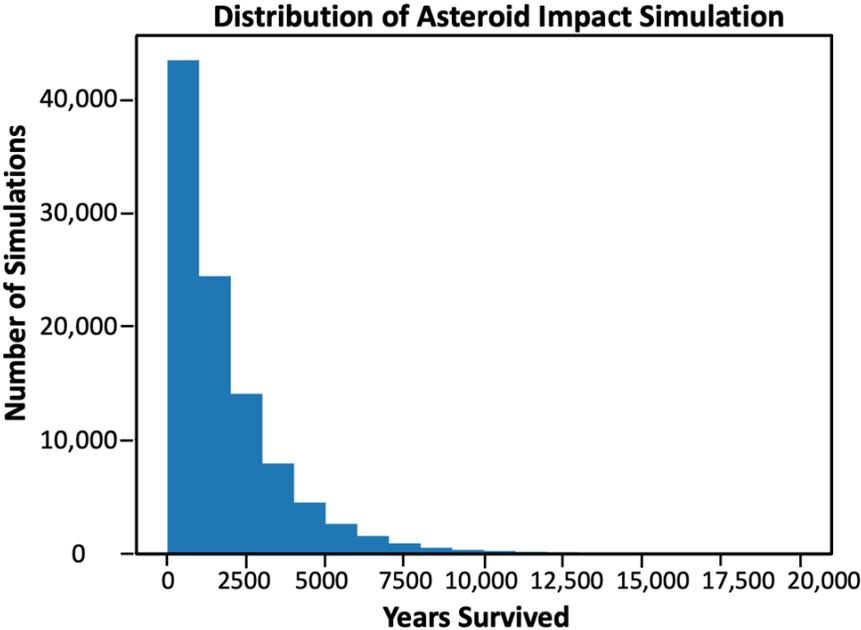

**Figure 15:** Distribution plot for asteroid impact simulation.

Alongside analyzing the existential risks from nuclear war, climate change, asteroid impacts and even artificial intelligence, pandemics are a major consideration. Pandemics not only claim the lives of countless numbers of innocent individuals, but also threaten the global economy and social infrastructure as acutely exemplified by COVID-19. The pandemic simulation is an attempt to predict possible Influenza-type events in the near future based on their having occupied a greater portion of the history of global pandemics. It is worth pointing out that unlike some other major threats to human survival, pandemics do not necessarily escalate very rapidly, but rather are more likely to experience a modest beginning as local outbreaks in any corner of the world, then increase in expanse to epidemics by spreading across a large region or a country, finally transforming into widespread global pandemics if not halted. This sheds light on an important fact - it is indeed possible to prevent epidemics from becoming pandemics by taking the right actions at the right times. Therefore, our simulation only implies the possibility of this scenario but does not guarantee it. The model utilizes basic reproduction numbers and attack rates as its basis since these are the two major parameters associated with epidemiological studies. For the inclusion of real-world data, the 1918 Influenza pandemic was chosen owing to its disastrous effects on human life and global turmoil it precipitated. From there, the attack rate was computed while also considering the threshold over which the possibility of a new pandemic emerges. Being a probabilistic measure, we have used the attack rate as a means of comparison with randomly generated probability for 100,000 epochs in order to compute the prediction results.

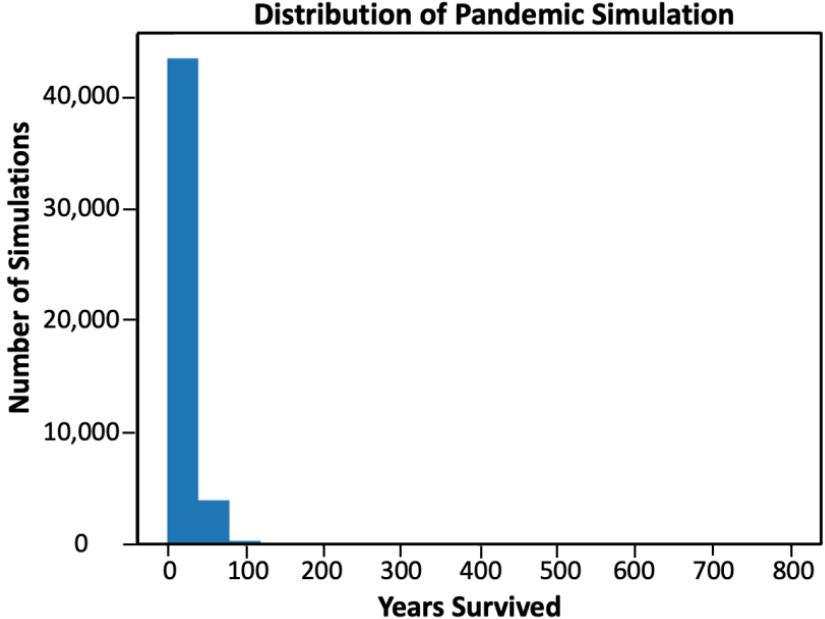

Figure 16: Distribution plot for pandemic simulation.

Based on the distribution plot in Figure 16, it can be visualized that most of the simulations fail to show survival ranging beyond the century mark. This prognosis prompts the notion of a binary choice: either wait for the next deadly epidemic to emerge into a pandemic, taking no impactful preparatory actions, or we take a proactive approach such as creation of a dedicated GERM team and rapidly improve medical infrastructure uniformly across all nations. Only through application of advanced medical technology and global cooperation – i.e., sound policies that safeguard our survival, can there be assurance of long-term sustainability of humanity on this *pale blue dot of a home* we call Earth.

## 5. Discussions and Implications

Throughout history we have repeatedly come to the realization that we are far from well-equipped to tackle apocalyptic scenarios – but nonetheless have survived to this point. Measures going forward have been taken as lessons from our destructive past - vaccines for viruses, laws and alliances to prevent racially motivated genocides, general leaps and bounds in infrastructure and computational ability, etc., improving the general quality of life for humanity. And yet, backward steps continue to pervade: faulty or poorly distributed resources in healthcare and polluted social climates that allow for camps such as those the Uhygars occupy in China. These are some of the few drastic impulses that, if left unchecked, may eventually lead to our destruction. Cultural philosophies driven by extremes in personal greed or soul-crushing collectivism give rise to self-reinforcing cause and effect cycles of societal destruction. It is difficult to see beyond the muddle of our daily lives to a theorized and modeled existential disaster that may threaten our species as a whole. But see we must.

That we are able to analytically contemplate our own demise, pragmatically putting numbers to scenarios, constitutes a clear first step towards deriving workable solutions. Diagnosis precedes formulation and application of the right curatives, and while the human condition cannot be called well it is far from dire and the prognosis holds hope. As convincingly argued by noted Harvard psychologist Steven Pinker [30], recent decades have witnessed a greater proportion of humanity gaining access to clean water, adequate shelter, nutritional sustenance, electricity, personal liberty and basic healthcare than at any time in history. Taking Professor Pinker's contention further, the digital information revolution has brought the collective knowledge of 5200 years of human civilization to the fingertips of billions in what perhaps will ultimately prove to be the most transformational development of modern society. So many human minds are now equipped with the knowhow - and the computational power - to transform ideas into reality. Much mischief can and has been carried out as well with these new tools, but the potential for the greater good to emerge victorious is unmistakably present.

The same large brains which helped our distant ancestors survive on the plains of Africa to eventually spread across the globe and build an ever more complex technological civilization holds both the cause and solution to our predicament. Ascending to mastery of our home world empowers us to control our own destiny and thus avoid the Great Filter – particularly those of our own making. Even in the highly unlikely occurrence of mass extinction events not brought about by our hand, such as a massive asteroid impact or super volcano eruptions, these needn't result in wiping out all of humanity if we have already established robust, self-sustaining and diverse colonies off-world. The human mind is the very pinnacle of evolution on Earth and the seat of technological invention and innovation. When properly harnessed for good ends, it will again prove to be our strongest ally as we face down the threats discussed in this study, defying the Great Filter and offering up to the cosmos Earth's response to Fermi's Paradox.

**Acknowledgements:** Authors JHJ and KAF are supported by the Jet Propulsion Laboratory, California Institute of Technology, under contract with NASA. We also acknowledge the supports by Carnegie Mellon University, Pittsburgh, Vivekananda Mission High School, Beijing Normal University, Sage Hill School, University of California Santa Barbara, and Radboud University.

**Data availability:** All data and software used for this study can be downloaded from https://github.com/Prithwis-2023/Group-1. For additional questions regarding the data sharing, please contact the corresponding author at Jonathan.H.Jiang@jpl.nasa.gov.

# References

1. Snyder-Beattie, A. E., Ord, T., & Bonsall, M. B., An upper bound for the background rate of human extinction, *Scientific Reports, 9.*, 2019.
2. Cai, X., J.H. Jiang, K. Fahy, Y. Yung, A Statistical Estimation of the Occurrence of Extraterrestrial Intelligence in the Milky Way Galaxy, *Galaxies*, 9, 1, 2021.
3. Hart, Michael H., Explanation for the Absence of Extraterrestrials on Earth, *Quarterly Journal of the Royal Astronomical Society.* 16: 128–135, 1975.
4. Sotos J.G., Biotechnology and the lifetime of technical civilizations. *International Journal of Astrobiology*, Vol 19, Issue 5, 2019.
5. Federation of American Scientists, *Status of World Nuclear Forces*. Federation Of American Scientists, from https://fas.org/issues/nuclear-weapons/status-world-nuclear-forces/ last accessed on August 28, 2022.
6. Dvorsky, G., *Incoming star could spawn swarms of comets when it passes our sun*, published December 23, 2016, from https://gizmodo.com/incoming-star-could-spawn-swarms-of-comets-when-it-pass-1790406698 last accessed on August 28, 2022.
7. Berski, Filip, Dybczyński, Piotr A., Gliese 710 will pass the Sun even closer, *Astronomy & Astrophysics,* 595, L10, 2016.
8. Copeland, B., Artificial intelligence. *Encyclopedia Britannica*, https://www.britannica.com/technology/artificial-intelligence, last updated August 24, 2022.
9. Dvorsky G., How an artificial superintelligence might actually destroy humanity, *Gizmodo*, published on May 26, 2021 at https://gizmodo.com/how-an-artificial-superintelligence-might-actually-dest-1846968207.
10. Brain performance in FLOPS, https://aiimpacts.org/brain-performance-in-flops/ last accessed on August 28, 2022.
11. Brain performance in TEPS, https://aiimpacts.org/brain-performance-in-teps/ last accessed on August 28, 2022.
12. Leprince-Ringuet, D., (2020, The world's fastest supercomputers are still getting faster, but it's taking them longer, *ZDNet*, published on November 17, 2020 at https://www.zdnet.com/article/the-worlds-fastest-supercomputers-are-still-getting-faster-but-its-taking-them-longer/.
13. Artificial general intelligence, *Wikipedia, The Free Encyclopedia*, https://en.wikipedia.org/w/index.php?title=Artificial_general_intelligence&oldid=1083529650, last accessed on August 28, 2022.
14. Gates, B., How to prevent the next pandemic, published on May 3, 2022, publisher: *Alfred A. Knopf*, New York, Toronto, 2022.
15. Reed, C., Biggerstaff, M., Finelli, L., Koonin, L. M., Beauvais, D., Uzicanin, A., Plummer, A., Bresee, J., Redd, S. C., & Jernigan, D. B., Novel framework for assessing epidemiologic effects of influenza epidemics and pandemics. *Emerging Infectious Diseases*, *19*(1), 85–91, 2013, https://doi.org/10.3201/eid1901.120124
16. Bruce W. Clements, Julie Ann P. Casani,16 - Pandemic Influenza, Editor(s): Bruce W. Clements, Julie Ann P. Casani, Disasters and Public Health (Second Edition), *Butterworth-Heinemann*, 2016, Pages 385-410, ISBN 9780128019801, https://doi.org/10.1016/B978-0-12-801980-1.00016-7.

17. Chapter 25 - Vaccine-Preventable Diseases, Editor(s): Dennis K. Flaherty, Immunology for Pharmacy, Mosby, 2012, Pages 197-213, ISBN 9780323069472, https://doi.org/10.1016/B978-0-323-06947-2.10025-2

18. Julien Arino, Chris Bauch, Fred Brauer, S. Michelle Driedger, Amy L. Greer, S.M. Moghadas, Nick J. Pizzi, Beate Sander, Ashleigh Tuite, P. van den Driessche, James Watmough, Jianhong Wu, Ping Yan. Pandemic influenza: Modelling and public health perspectives, *Mathematical Biosciences and Engineering*, 2011, 8(1): 1-20. doi:10.3934/mbe.2011.8.1

19. Biggerstaff, M., Cauchemez, S., Reed, C. *et al.* Estimates of the reproduction number for seasonal, pandemic, and zoonotic influenza: a systematic review of the literature. *BMC Infect Dis* **14,** 480 (2014). https://doi.org/10.1186/1471-2334-14-480

20. Maragakis, L. L., Covid-19 vs. the flu. Johns Hopkins Medicine, last updated on https://www.hopkinsmedicine.org/health/conditions-and-diseases/coronavirus/coronavirus-disease-2019-vs-the-flu last accessed on August 28, 2022.

21. Centers for Disease Control and Prevention, Lesson 3: Measures of Risk, https://www.cdc.gov/csels/dsepd/ss1978/lesson3/section2.html last accessed on August 28, 2022.

22. Effects of nuclear explosions, *Wikipedia,* https://en.wikipedia.org/wiki/Effects_of_nuclear_explosions last accessed on August 28, 2022.

23. The World Bank. (n.d.). Urban land area (sq. km), from https://data.worldbank.org.cn/indicator/AG.LND.TOTL.UR.K2 last accessed on August 28, 2022.

24. R. P. Turco, O. B. Toon, T. P. Ackerman, J. B. Pollack and Carl Sagan, Nuclear Winter: Global Consequences of Multiple Nuclear Explosions, Science, Vol. 222, No. 4630, pp. 1283- 1292, 1983.

25. Thompson, Starley L., and Stephen H. Schneider, Nuclear Winter Reappraised, *Council on Foreign Affairs*, Vol. 64, No. 5, 981–1005, 1986. https://doi.org/10.2307/20042777.

26. Hans M. Kristensen, Matt Korda and Robert Norris Estimated Global Nuclear Warhead Inventories 1945-2022, *Federation of American*, 2022, on https://fas.org/issues/nuclear-weapons/status-world-nuclear-forces/ last accessed on August 28, 2022.

27. Jiang, J.H.; Feng, F.; Rosen, F.; Fahy, K.A.; Das, P.; Obacz, P.; Zhang, A.; Zhu, Z.-H. Avoiding the Great Filter: Predicting the Timeline for Humanity to Reach Kardashev Type I Civilization. *Galaxies*,10, 68, 2022.

28. Bohn, K., published on March 1, 2022 at https://www.psu.edu/news/research/story/humans-cant-endure-temperatures-and-humidities-high-previously-thought/ last accessed on August 28, 2022.

29. J. R. Powell, The Quantum Limit to Moore's Law, *Proceedings of the IEEE*, vol. 96, no. 8, pp. 1247-1248, Aug. 2008, doi:10.1109/JPROC.2008.925411.

30. Pinker, S.A., Enlightenment Now: The case for reason, science, humanism, and progress, *Penguin Books/Penguin Random House* LLC, USA, (2018), ISBN 978-0-525-42757-5.

31. GISS Surface Temperature Analysis, https://data.giss.nasa.gov/gistemp/graphs/graph_data/Global_Mean_Estimates_based_on_Land_and_Ocean_Data/graph.txt last accessed on August 28, 2022.

32. AI and Compute, https://openai.com/blog/ai-and-compute/ last accessed on September 18, 2022.

33. Measuring compute capacity: a critical step to capturing AI's full economic potential, https://oecd.ai/en/wonk/ai-compute-capacity, last accessed on September 18, 2022.
34. The Great Filter: Are We Almost Past It?, https://mason.gmu.edu/~rhanson/greatfilter.html, last accessed on September 20, 2022.